\documentclass[debug,overfull]{epl}

\title{Quantum Correction in Exact Quantization Rules}

\author{Zhong-Qi Ma\inst{1}\thanks{Electronic address:
mazq@mail.ihep.ac.cn} \and Bo-Wei Xu\inst{2}\thanks{Electronic
address: bwxu@sjtu.edu.cn}}

\institute{ \inst{1} Institute of High Energy Physics, P. O. Box
918(4), Beijing 100049, China\\
\inst{2} Department of Physics, Shanghai Jiaotong University,
Shanghai 200030, China}


\begin{document}
\maketitle

\date{}

\vspace{5mm}

\begin{abstract}

An exact quantization rule for the Schr\"{o}dinger equation is
presented. In the exact quantization rule, in addition to $N\pi$,
there is an integral term, called the quantum correction. For the
exactly solvable systems we find that the quantum correction is an
invariant, independent of the number of nodes in the wave
function. In those systems, the energy levels of all the bound
states can be easily calculated from the exact quantization rule
and the solution for the ground state, which can be obtained by
solving the Riccati equation. With this new method, we
re-calculate the energy levels for the one-dimensional systems
with a finite square well, with the Morse potential, with the
symmetric and asymmetric Rosen-Morse potentials, and with the
first and the second P\"{o}schl-Teller potentials, for the
harmonic oscillators both in one dimension and in three
dimensions, and for the hydrogen atom.

\end{abstract}

In the development of quantum mechanics, the Bohr-Sommerfeld
quantization rules of the old quantum theory \cite{sch} occupy a
position intermediate between classical and quantum mechanics. The
WKB approximation \cite{wen,kra,bri} is a method for the
approximate treatment of the Schr\"{o}dinger wave function with
another quantization rule \cite{sch}. Both quantization rules are
approximate. We are going to derive an exact quantization rule for
the one-dimensional Schr\"{o}dinger equation:
$$\displaystyle {d^{2}\over dx^{2}}\psi(x)=-\displaystyle {2M\over
\hbar^{2}} \left[E-V(x)\right]\psi(x), \eqno (1) $$

\noindent where $M$ is the mass of the particle, and the potential
$V(x)$ is a piecewise continuous real function of $x$ satisfying
for definiteness
$$\begin{array}{ll}
V(x)>E, ~~~~~~~~~~&-\infty < x < x_{A}~~{\rm or}~~x_{B}< x < \infty,\\
V(x)=E, ~~~~&x=x_{A}~~{\rm or}~~x=x_{B},\\
V(x)<E, ~~~~&x_{A}<x<x_{B}, \end{array} \eqno (2) $$

\noindent where $x_{A}$ and $x_{B}$ are two turning points.
Between two turning points, the momentum is
$k(x)=\sqrt{2M\left[E-V(x)\right]}/\hbar$.

Yang pointed out in a talk on monopole: ``For the Sturm-Liouville
problem, the fundamental trick is the definition of a phase angle
which is monotonic with respect to the energy." \cite{yan}$~$ This
phase angle is the logarithmic derivative
$\phi(x)=\psi(x)^{-1}d\psi(x)/dx$ of the wave function $\psi(x)$.
Due to the Sturm-Liouville theorem, $\phi(x)$ at any given point
$x=x_{0}$ is monotonic with respect to the energy. From the
Schr\"{o}dinger equation (1), $\phi(x)$ satisfies the Riccati
equation:
$$\displaystyle {d\over dx}\phi(x)=-\displaystyle {2M\over \hbar^{2}}
\left[E-V(x)\right]-\phi(x)^{2}. \eqno (3) $$

\noindent It shows that $\phi(x)$ decreases monotonically with
respect to $x$ between two turning points where $E\geq V(x)$. Note
that as $x$ increases across a node of the wave function $\psi(x)$
where $E\geq V(x)$, $\phi(x)$ decreases to $-\infty$, jumps to
$+\infty$, and then, decreases again. The Riccati equation is a
differential equation of the first order so that it is much easier
to find a special solution from the Riccati equation than from the
Schr\"{o}dinger equation.

Letting $\tan \theta(x)=k(x)/\phi(x)$, we have $\theta(x)={\rm
Arctan}\left[k(x)/\phi(x)\right]+n\pi$, where Arctan $\beta$
denotes the principle value of the inverse tangent function,
$-\pi/2< {\rm Arctan}~ \beta \leq \pi/2$, and $n$ increases by one
as $x$ increases across a node of $\phi(x)$ where $E\geq V(x)$.
Then, we have
$$\displaystyle \int_{x_{A}}^{x_{B}}\displaystyle {d\theta(x)\over dx}dx
= N\pi -\displaystyle \lim_{x\rightarrow x_{A}+}~{\rm
Arctan}\left(\displaystyle {k(x)\over \phi(x_{A})}\right)
+\displaystyle \lim_{x\rightarrow x_{B}-}~{\rm
Arctan}\left(\displaystyle{ k(x)\over \phi(x_{B})}\right), \eqno
(4) $$

\noindent where $N$ is the number of nodes of $\phi(x)$, and two
terms with limit are vanishing if the potential $V(x)$ is
continuous at the turning points. From Eq. (3) we have
$$\displaystyle {d\theta(x)\over dx}=k(x)-\phi(x)\left[\displaystyle
{dk(x)\over dx}\right] \left[\displaystyle {d\phi(x)\over
dx}\right]^{-1},~~~~~~E\geq V(x). \eqno (5) $$

\noindent Integrating both sides of Eq. (5) over the interval
between two turning points, we obtain the quantization rule
without any approximation:
$$\displaystyle
\int_{x_{A}}^{x_{B}}k(x)dx = N\pi+\displaystyle
\int_{x_{A}}^{x_{B}} \phi(x)\left[\displaystyle {dk(x)\over
dx}\right] \left[\displaystyle {d\phi(x)\over dx}\right]^{-1} dx.
\eqno (6)$$

\noindent Since $\phi(x_{A})> 0$, $\phi(x_{B})< 0$, and $\phi(x)$
decreases monotonically in the region $x_{A}<x<x_{B}$, the number
$N$ of nodes of $\phi(x)$ in that region is larger by one than the
number of nodes of the wave solution $\psi(x)$. Due to the Sturm
theorem, the number $(N-1)$ of nodes of $\psi(x)$ increases as the
energy $E$ increases.

Generalize the quantization rule to the three-dimensional
Schr\"{o}dinger equation with a spherically symmetric potential.
After separation of the angular part of the wave function,
$\psi({\bf r})=r^{-1}R(r)Y^{\ell}_{m}(\theta,\varphi)$, the radial
equation of the Schr\"{o}dinger equation is
$$\displaystyle {d^{2}R(r)\over dr^{2}}=-\displaystyle {2M\over
\hbar^{2}}\left[E-U(r)\right]R(r),~~~~~~ U(r)= \displaystyle
{\ell(\ell+1)\hbar^{2}\over 2M r^{2}}+V(r). \eqno (7) $$

\noindent Since Eq. (7) is similar to Eq. (1), the quantization
rule (6) is generalized to the three-dimensional Schr\"{o}dinger
equation with a spherically symmetric potential
$$\displaystyle \int_{r_{A}}^{r_{B}}k(r)dr = N\pi+\displaystyle
\int_{r_{A}}^{r_{B}} \phi(r)\left[\displaystyle {dk(r)\over
dr}\right] \left[\displaystyle {d\phi(r)\over dr}\right]^{-1} dr.
\eqno (8)$$

The quantization rule (6) [or (8)] is proved without any
approximation, so that it is exact. The first term $N\pi$ in the
quantization rule relates to the contribution from the nodes of
the wave function, and the second term is called the quantum
correction. We find that the quantum correction is independent of
the number of nodes of the wave function for the exactly solvable
systems.  In those systems, one is able to calculate easily the
energy levels of all bound states from the exact quantization rule
and the solution of the ground state, which can be calculated
directly from the Riccati equation. With this new method, we are
going to re-calculate the energy levels of those systems.

The one-dimensional system with a finite square well, where
$V(x)=V_{A}$ when $x\leq -\pi$, $V(x)=V_{B}$ when $x\geq \pi$, and
$V(x)=0$ when $-\pi<x<\pi$, is a typical example in the course of
quantum mechanics. However, the solutions can be obtained with the
exact quantization rule much simpler than with the standard
method. Since $k(x)$ is constant between two turning points, the
quantum correction in the quantization rule is vanishing. For the
finite square well, two terms with limit in Eq. (4) has to be
included in the quantization rule (6) because the potential jumps
at the turning points. The energy levels of the system is directly
given in the quantization rule (6) and (4) \cite{mes}:
$$2\pi k_{N}=N\pi-{\rm Arctan}\left(\displaystyle {k_{N}\over \phi_{N}(x_{A})}\right)
+{\rm Arctan}\left(\displaystyle{ k_{N}\over
\phi_{N}(x_{B})}\right), \eqno (9) $$

\noindent where $\phi_{N}(x_{A})=\sqrt{2M(V_{A}-E_{N})}/\hbar$,
$\phi_{N}(x_{B})=-\sqrt{2M(V_{B}-E_{N})}/\hbar$. When
$V_{A}=V_{B}\rightarrow \infty$, we have $k_{N}=N/2$ \cite{sch}.

The potential for a one-dimensional harmonic oscillator is $V(x)=M
\omega^{2}x^{2}/2$. The turning points are solved to be
$x_{B}=-x_{A}=\alpha^{-1}\sqrt{2E_{n}/(\hbar \omega)}$, where
$\alpha=\sqrt{M \omega/\hbar}$, and $(n+1)$ is the number of nodes
of $\phi_{n}(x)$. The momentum between two turning points is
$k_{n}(x)=\alpha^{2} \left[ \left(x-x_{A}\right)
\left(x_{B}-x\right)\right]^{1/2}$. The solution with one node and
no pole only has the form as $\phi_{0}(x)=-C x$ where $C>0$ due to
the monotonic property. Substituting it into the Riccati equation
(3) with the potential of a harmonic oscillator, we find
$\phi_{0}(x)=-\alpha^{2} x$ with $E_{0}=\hbar\omega/2$. Evidently,
$\phi_{0}(x)$ is negative when $x\rightarrow \infty$ and is
positive when $x\rightarrow -\infty$, so that the solution
satisfies the physically admissible boundary conditions. All the
solutions in the following examples have the similar behavior. Two
integrals in the exact quantization rule (6) are calculated to be
$$\displaystyle
\int_{x_{A}}^{x_{B}} \phi_{0}(x)\left[\displaystyle
{dk_{0}(x)\over dx}\right] \left[\displaystyle {d\phi_{0}(x)\over
dx}\right]^{-1} dx=-\pi/2. \eqno (10) $$
$$\displaystyle \int_{x_{A}}^{x_{B}}k_{n}(x)dx =E_{n}\pi/(\hbar \omega).
\eqno (11) $$

The quantization rule (6) coincides with the quantization rule in
WKB approximation:
$$\displaystyle \int_{x_{A}}^{x_{B}}k_{n}(x)dx =(n+1/2)\pi. \eqno (12) $$

\noindent The energy levels for the one-dimensional harmonic
oscillator are \cite{sch}:
$$E_{n}=(n+1/2)\hbar \omega. \eqno (13) $$

The one-dimensional Morse potential is
$V(x)=D\left(e^{-2x/a}-2e^{-x/a}\right)=Dy(y-2)$, where
$y=e^{-x/a}$. The turning points are
$y_{A}=e^{-x_{A}/a}=1-\sqrt{1+E_{n}/D}$ and
$y_{B}=e^{-x_{B}/a}=1+\sqrt{1+E_{n}/D}$, where $(n+1)$ denotes the
number of nodes of the logarithmic derivative $\phi_{n}(x)$. The
momentum between two turning points is $k_{n}(x)= \left[(2M
D/\hbar^{2})(y-y_{A})(y_{B}-y)\right]^{1/2}$. The solution with
only one node has to be $\phi_{0}(x)=C_{1}y+C_{2}$ with $C_{1}>0$.
Substituting it into the Riccati equation (3) with the Morse
potential, we find $\phi_{0}(x)=\sqrt{2M D}(y-1)/ \hbar+1/(2a)$
with $E_{0}=-[\sqrt{D}-\hbar / (a \sqrt{8M})]^{2}$. Two integrals
in the exact quantization rule (6) are calculated to be

$$\displaystyle
\int_{x_{A}}^{x_{B}} \phi_{0}(x)\left[\displaystyle
{dk_{0}(x)\over dx}\right] \left[\displaystyle {d\phi_{0}(x)\over
dx}\right]^{-1} dx=-\pi/2, \eqno (14) $$
$$\displaystyle \int_{x_{A}}^{x_{B}}k_{n}(x)dx
=\displaystyle {a \sqrt{2M}\over \hbar
}\left[\sqrt{D}-\sqrt{-E_{n}}\right] \pi.  \eqno (15) $$

\noindent The quantization rule is $\int_{x_{A}}^{x_{B}}k_{n}(x)dx
= (n+1/2)\pi$. Thus, the energy levels for the one-dimensional
system with the Morse potential are \cite{lan}:
$$E_{n}=-\left[\sqrt{D}-\displaystyle {(2n+1)\hbar \over 2a
\sqrt{2M}}\right]^{2} . \eqno (16)$$

The asymmetric Rosen-Morse potential \cite{lan} in one dimension
is $V(x)=-U_{0}{\rm sech}^{2}(x/a)+U_{1}\tanh(x/a)$, where $0\leq
U_{1}<2U_{0}$. If $U_{1}=0$, $V(x)$ is called the symmetric
Rosen-Morse potential. Let $y=\tanh (x/a)$, $y_{A}=\tanh
(x_{A}/a)$, and $y_{B}=\tanh (x_{B}/a)$, where $x_{A}$ and $x_{B}$
are two turning points satisfying $V(x_{A})=V(x_{B})=E_{n}$. We
have
$$y_{A}=-\displaystyle {U_{1}\over 2U_{0}}-
\sqrt{\left(\displaystyle {U_{1}\over
2U_{0}}\right)^{2}+\displaystyle {E_{n}\over U_{0}}+1},~~~~
y_{B}=-\displaystyle {U_{1}\over 2U_{0}}+
\sqrt{\left(\displaystyle {U_{1}\over
2U_{0}}\right)^{2}+\displaystyle {E_{n}\over U_{0}}+1}, $$

\noindent where $(n+1)$ is the number of nodes of the logarithm
derivative $\phi_{n}(x)$. The momentum between two turning points
is $k_{n}(x)= \sqrt{2M U_{0}/\hbar^{2}} [ (y-y_{A})
(y_{B}-y)]^{1/2}$. From the Riccati equation (3) with the
asymmetric Rosen-Morse potential we obtain the solution with one
node
$$\phi_{0}(x)=-\displaystyle {1 \over
2a}\left\{\left(1+\displaystyle {8a^{2}M U_{0} \over
\hbar^{2}}\right)^{1/2}-1\right\}y-\displaystyle {M U_{1} \over
\hbar^{2}C}~~~~~{\rm with}~~E_{0}=-\displaystyle { \hbar^{2} C^{2}
\over 2M}-\displaystyle {M U_{1}^{2}\over 2\hbar^{2}C^{2}}. $$

Now, two integrals in the quantization rule (6) are calculated to
be
$$\displaystyle
\int_{x_{A}}^{x_{B}} \phi_{0}(x)\left[\displaystyle
{dk_{0}(x)\over dx}\right] \left[\displaystyle {d\phi_{0}(x)\over
dx}\right]^{-1} dx=\displaystyle {a\pi \sqrt{2M U_{0}} \over
\hbar}\left[1-\displaystyle {\sqrt{2M U_{0}} \over \hbar
C}\right]. \eqno (17)$$
$$\displaystyle \int_{x_{A}}^{x_{B}}k_{n}(x)dx
=\displaystyle {a\pi\over 2}\displaystyle {\sqrt{2M U_{0}}\over
\hbar}\left[ 2-\sqrt{\displaystyle {-E_{n}-U_{1}\over U_{0}}}
-\sqrt{\displaystyle {-E_{n}+U_{1}\over U_{0}}}\right].
 \eqno (18)$$

\noindent The quantization rule (6) becomes:
$$\displaystyle {1\over 2}\sqrt{\displaystyle {-E_{n}-U_{1}\over
U_{0}}} +\displaystyle {1\over 2}\sqrt{\displaystyle
{-E_{n}+U_{1}\over U_{0}}} =-\displaystyle {(n+1)\hbar \over
a\sqrt{2M U_{0}}}+\displaystyle {\sqrt{2M U_{0}} \over \hbar
C}=\displaystyle {(Ca-n)\hbar \over a\sqrt{2M U_{0}}}. \eqno
(19)$$

\noindent Thus, the energy $E_{n}$ is \cite{lan}
$$-E_{n}=\displaystyle {
\hbar^{2} (C-n/a)^{2} \over 2M}+\displaystyle {M U_{1}^{2}\over
2\hbar^{2}(C-n/a)^{2}}.  \eqno (20) $$

\noindent The condition of existence for the bound state whose
wave function has $n$ nodes is
$U_{1}<\hbar^{2}(C-n/a)^{2}/M<2U_{0}$. When $U_{1}=0$, the
asymmetric Rosen-Morse potential becomes the symmetric one. The
energy levels (20) with $U_{1}=0$ hold for the symmetric
Rosen-Morse potential \cite{lan}.

The first P\"{o}schl-Teller potential \cite{lan} in one dimension
is
$$V(x)=\displaystyle {\hbar^{2} \over
2Ma^{2}}\left[\displaystyle { \mu(\mu-1) \over
\sin^{2}(x/a)}+\displaystyle { \lambda(\lambda-1) \over
\cos^{2}(x/a)}\right],~~~~~~0< x < \displaystyle {a \pi \over
2},\eqno (21) $$

\noindent where $\mu$ and $\lambda$ are constant greater than one.
The potential $V(x)$ tends to infinity as $x$ tends to $0$ or
$a\pi/2$. Let $y=\tan^{2}(x/a)$, $y_{A}=\tan^{2}(x_{A}/a)$, and
$y_{B}=\tan^{2}(x_{B}/a)$, where $x_{A}$ and $x_{B}$ are two
turning points satisfying $V(x_{A})=V(x_{B})=E_{n}$. We have
$$y_{A}+y_{B}=\displaystyle
{2Ma^{2}E_{n}/\hbar^{2}-\mu(\mu-1)-\lambda(\lambda-1)\over
\lambda(\lambda-1)},~~~~~~ y_{A}y_{B}=\displaystyle
{\mu(\mu-1)\over \lambda(\lambda-1)},   $$

\noindent where $(n+1)$ denotes the number of nodes of the
logarithm derivative $\phi_{n}(x)$. The momentum $k_{n}(x)$
between two turning points is $ k_{n}(x) =[\lambda(\lambda-1)
(y-y_{A})(y_{B}-y)/y]^{1/2}/a$ when  $E\geq V(x)$. The solution of
the Riccati equation with only one node is $\phi_{0}(x)=-\lambda
y^{1/2}/ a+ \mu y^{-1/2}/a$ with
$E_{0}=\hbar^{2}(\mu+\lambda)^{2}/(2Ma^{2})$.

Two integrals in the quantization rule (6) are calculated to be
$$\displaystyle
\int_{x_{A}}^{x_{B}} \phi_{0}(x)\left[\displaystyle
{dk_{0}(x)\over dx}\right] \left[\displaystyle {d\phi_{0}(x)\over
dx}\right]^{-1} dx=\displaystyle {\pi  \over
2}\left[(\mu+\lambda-2)-\sqrt{\mu(\mu-1)}-\sqrt{\lambda(\lambda-1)}\right].
\eqno (22)$$
$$\displaystyle \int_{x_{A}}^{x_{B}}k_{n}(x)dx =\displaystyle {\pi
\over 2} \left[\displaystyle {a\sqrt{2ME_{n}}\over \hbar}
-\sqrt{\mu(\mu-1)}-\sqrt{\lambda(\lambda-1)}\right]. \eqno (23)$$

\noindent  The quantization rule (6) becomes $a\sqrt{2ME_{n}}/
\hbar=2(n+1)+(\mu+\lambda-2)$, from which we obtain the energy
$E_{n}$  \cite{lan}
$$E_{n}=\displaystyle {\hbar^{2}(\mu+\lambda+2n)^{2}\over
2Ma^{2}}.  \eqno (24) $$

The second P\"{o}schl-Teller potential \cite{lan} in one dimension
is
$$V(x)=\displaystyle {\hbar^{2} \over
2Ma^{2}}\left[\displaystyle { \mu(\mu-1) \over
\sinh^{2}(x/a)}-\displaystyle { \lambda(\lambda+1) \over
\cosh^{2}(x/a)}\right], \eqno (25) $$

\noindent where $\lambda>\mu-1>0$. The potential $V(x)$ tends to
infinity at $x=0$. Let $y=\tanh^{2}(x/a)$, $
y_{A}=\tanh^{2}(x_{A}/a)$, and $y_{B}=\tanh^{2}(x_{B}/a)$ where
$x_{A}$ and $x_{B}$ are two turning points where
$V(x_{A})=V(x_{B})=E_{n}$. We have
$$y_{A}+y_{B}=\displaystyle
{2Ma^{2}E_{n}/\hbar^{2}+\mu(\mu-1)+\lambda(\lambda+1)\over
\lambda(\lambda+1)},~~~~~~ y_{A}y_{B}= \displaystyle
{\mu(\mu-1)\over \lambda(\lambda+1)},  $$

\noindent where $(n+1)$ denotes the number of nodes of the
logarithm derivative $\phi_{n}(x)$. The momentum $k_{n}(x)$
between two turning points is
$k_{n}(x)=[\lambda(\lambda+1)(y-y_{A})(y_{B}-y)/y]^{1/2}/a$, when
$ E\geq V(x)$. The solution of the Riccati equation with only one
node is $\phi_{0}(x)=-\lambda y^{1/2}/a+ \mu y^{-1/2}/a$ with
$E_{0}=-\hbar^{2}(\lambda-\mu)^{2}/(2Ma^{2})$.

Two integrals in the quantization rule (6) are calculated to be
$$\displaystyle
\int_{x_{A}}^{x_{B}} \phi_{0}(x)\left[\displaystyle
{dk_{0}(x)\over dx}\right] \left[\displaystyle {d\phi_{0}(x)\over
dx}\right]^{-1} dx=\displaystyle {\pi  \over
2}\left[(\mu-\lambda-2)-\sqrt{\mu(\mu-1)}+\sqrt{\lambda(\lambda+1)}\right].
\eqno (26)$$
$$\displaystyle \int_{x_{A}}^{x_{B}}k_{n}(x)dx =\displaystyle {\pi
\over 2} \left[\displaystyle {a\sqrt{-2ME_{n}}\over \hbar}
-\sqrt{\mu(\mu-1)}+\sqrt{\lambda(\lambda+1)} \right]. \eqno (27)$$

\noindent  The quantization rule (6) reads $a\sqrt{-2ME_{n}}/
\hbar=2(n+1)+(\mu-\lambda-2)$, and the energy $E_{n}$ is
\cite{lan}
$$E_{n}=-\displaystyle
{\hbar^{2}(\lambda-\mu-2n)^{2}\over 2Ma^{2}},~~~~~~0\leq n <
(\lambda-\mu)/2. \eqno (28) $$

The effective potential for the three-dimensional harmonic
oscillator is  $U_{\ell}(r)=\ell(\ell+1)\hbar^{2}/(2 M r^{2})+ M
\omega^{2}r^{2}/2$. The turning points are
$r_{A}=\alpha^{-1}\{[E_{n\ell}/(\hbar \omega)]-[(E_{n\ell}/ (\hbar
\omega))^{2}-\ell(\ell+1)]^{1/2}\}^{1/2}$ and
$r_{B}=\alpha^{-1}\{[E_{n\ell}/(\hbar \omega)]+[(E_{n\ell}/ (\hbar
\omega))^{2}-\ell(\ell+1)]^{1/2}\}^{1/2}$, where $\alpha=\sqrt{M
\omega/\hbar}$, $(n-\ell)$ is a non-negative even integer, and
$(n-\ell+2)/2$ is the number of nodes of the logarithmic
derivative $\phi_{n\ell}(r)$. The momentum between two turning
points is $k_{n\ell}(r)=(\alpha^{2}/r)[(r^{2}-r_{A}^{2})
(r_{B}^{2}-r^{2})]^{1/2}$. The solution with only one node has to
be $\phi_{n\ell}(r)=C_{1}r^{-1}+C_{2}r$ where $C_{1}>0$ and
$C_{2}<0$. Substituting it into the Riccati equation (3) with the
harmonic oscillator potential, we find
$\phi_{n\ell}(r)=(\ell+1)r^{-1}-\alpha^{2} r$ with
$E_{n\ell}=\hbar \omega (\ell+3/2)$ and $n=\ell$.

Two integrals in the quantization rule (8) are calculated to be
$$\displaystyle
\int_{r_{A}}^{r_{B}} \phi_{n \ell}(r)\left(dk_{n
\ell}(r)/dr\right) \left[\displaystyle {d\phi_{n \ell}(r)\over
dr}\right]^{-1}
dr=\left[\ell-\sqrt{\ell(\ell+1)}-1/2\right]\pi/2,~~~~~~n=\ell.
\eqno (29)$$
$$\displaystyle \int_{r_{A}}^{r_{B}}k_{n\ell}(r)dr
=\left[\displaystyle {E_{n\ell} \over \hbar
\omega}-\sqrt{\ell(\ell+1)}\right]\pi/2.
 \eqno (30) $$

\noindent The quantization rule (8) becomes
$\int_{r_{A}}^{r_{B}}k_{n\ell}(r)dr
=[n-\sqrt{\ell(\ell+1)}+3/2]\pi/2$. Thus, the energy levels for
the three-dimensional harmonic oscillator are \cite{sch}:
$$E_{n\ell}=(n+3/2)\hbar \omega. \eqno (31) $$

The effective potential for the hydrogen atom is $U_{\ell}(r)=
\ell(\ell+1)\hbar^{2}/(2M r^{2}) - e^{2}/r$. When $\ell> 0$, the
turning points $r_{A}$ and $r_{B}$ satisfying
$U_{\ell}(r_{A})=U_{\ell}(r_{B})=E_{n\ell}$ are
$$\begin{array}{l}
r_{A}=\left(-2E_{n\ell}\right)^{-1}\left\{e^{2}-\left[e^{4}+
2\ell(\ell+1)\hbar^{2}E_{n\ell}/M\right]^{1/2}\right\},\\
r_{B}=\left(-2E_{n\ell}\right)^{-1}\left\{e^{2}+\left[e^{4}+
2\ell(\ell+1)\hbar^{2}E_{n\ell}/M\right]^{1/2}\right\},
\end{array} \eqno (32) $$

\noindent where we denote by $(n-\ell)$ the number of nodes of the
logarithmic derivative $\phi_{n\ell}(r)$. When $\ell=0$, we define
$r_{A}=0$ with $U(r_{A})=-e^{2}/r_{A}\sim -\infty$, and $r_{B}$ is
solved from $U(r_{B})=E_{n\ell}$. Thus, equation (32) still holds
for $\ell=0$. The momentum between two turning points is
$k_{n\ell}(r)=\left(\hbar r\right)^{-1}\left[-2M
E_{n\ell}(r-r_{A})(r_{B}-r)\right]^{1/2}$. For $\ell=0$, near the
origin we have $k_{n\ell}(r)\sim r^{-1/2}$, and $\phi_{n\ell}(r)
\sim r^{-1}$, so that the limit terms in Eq. (4) still vanish. The
 solution with only one node has the form as
$\phi_{n\ell}(r)=C_{1}r^{-1}+C_{2}$ where $C_{1}>0$. Combining it
with the Riccati equation (3) for the hydrogen atom, we find
$\phi_{n\ell}(r)=(\ell+1)/r-M e^{2}/[(\ell+1)\hbar^{2}]$ with
$E_{n\ell}=-M e^{4}/[ 2(\ell+1)^{2}\hbar^{2}]$ and $n=\ell+1$.

Two integrals in the quantization rule (8) are calculated to be
$$\displaystyle
\int_{r_{A}}^{r_{B}} \phi_{n\ell}(r)\left[\displaystyle
{dk_{n\ell}(r)\over dr}\right] \left[\displaystyle
{d\phi_{n\ell}(r)\over dr}\right]^{-1} \displaystyle
dr=\left[\ell-\sqrt{\ell(\ell+1)}\right]\pi,~~~~~~n=\ell+1. \eqno
(33)$$
$$\displaystyle \int_{r_{A}}^{r_{B}}k_{n\ell}(r)dr
=\left[\displaystyle {e^{2}\over \hbar}\sqrt{\displaystyle {M\over
-2E_{n\ell}}} -\sqrt{\ell(\ell+1)}\right] \pi.  \eqno (34) $$

\noindent The quantization rule (8) becomes
$\int_{r_{A}}^{r_{B}}k_{n\ell}(r)dr =[n-\sqrt{\ell(\ell+1)}]\pi$.
Thus, the energy levels for the hydrogen atom are \cite{sch}:
$$E_{n\ell}=-\displaystyle {M e^{4} \over
2n^{2}\hbar^{2}}. \eqno (35) $$

In this Letter we present an exact quantization rule (6) for the
one-dimensional Schr\"{o}dinger equation and (8) for the
three-dimensional Schr\"{o}dinger equation with a spherically
symmetric potential. We find that the quantum correction term is
independent of the number of nodes in the wave function for the
exactly solvable quantum systems.  In such cases, the energy
levels of the quantum system can be easily solved from the exact
quantization rule and the solution of the ground state calculated
directly from the Riccati equation. For the non-exactly solvable
systems, one can use the series form of the quantization rule for
numerical calculation \cite{ma}.

As is well known, the wave functions and the energy levels for the
exactly solvable systems can also be solved by the supersymmetry
in the recursive way \cite{coo}. The logarithmic derivatives
$\phi(x)$ are nothing but the superpotentials in the
supersymmetric quantum mechanics. The shape invariance of the
superpotentials seems to be related to the condition of the
quantum correction being an invariant. As far as the wave function
$\psi_{N}(x)$ with $(N-1)$ nodes is concerned, we prefer to
calculate its logarithmic derivatives $\phi_{N}(x)$ directly from
the Riccati equation (3). Although there are more or less
differences for different examples,  the solution for
$\phi_{N}(x)$ can be taken as a fraction, where the numerator is a
polynomial of order $N$ and the denominator is a polynomial of
order $(N-1)$, because $\phi_{N}(x)$ contains $N$ nodes and
$(N-1)$ poles. Without loss of generality, the coefficient of
$x^{N-1}$ in the denominator can be chosen to be one. Substituting
it into the Riccati equation (3), one obtains a coupled algebraic
equations of order two for the coefficients. Therefore,
$\phi_{N}(x)$ can be solved. We have solved some examples for the
lower excited states.

\vspace{5mm} \noindent {\bf ACKNOWLEDGMENTS}. The authors are
grateful to Professor C. N. Yang for stimulating discussions. This
work was supported by the National Natural Science Foundation of
China.

\end{document}